\documentstyle[aps,psfig]{revtex}
\begin{document}
\twocolumn[\hsize\textwidth\columnwidth\hsize\csname
@twocolumnfalse\endcsname
\title{Instability and `Sausage-String' Appearance\\
in Blood Vessels during High Blood Pressure}
\author{Preben Alstr{\o}m$^1$, Victor M. Egu\'{\i}luz$^2$,
Morten Colding-J{\o}rgensen$^3$,\\
Finn Gustafsson$^4$, and Niels-Henrik Holstein-Rathlou$^4$}
\address{$^1$CATS, The Niels Bohr Institute, DK-2100 Copenhagen, Denmark\\
$^2$Instituto Mediterr\'aneo de Estudios Avanzados IMEDEA (CSIC-UIB),
E-07071 Palma de Mallorca, Spain\\
$^3$Novo Nordisk, DK-2880 Bagsvaerd, Denmark\\
$^4$Department of Medical Physiology, The Panum Institute,
DK-2200 Copenhagen, Denmark}
\date{September 30, 1998}
\maketitle 
\begin{abstract}

A new Rayleigh-type instability is proposed to explain the `sausage-string'
pattern of alternating constrictions and dilatations formed in blood
vessels under influence of a vasoconstricting agent. Our theory
involves the nonlinear elasticity characteristics of the vessel wall,
and provides predictions for the conditions under which the
cylindrical form of a blood vessel becomes unstable.

\end{abstract}
]

High blood pressure can experimentally be induced by intravenous
infusion of a vasoconstricting agent like angiotensin II (regulates
the contraction of the smooth muscle cells surrounding the blood
vessel) \cite{bb,jg,g&g}. As the infusion is continued, a substantial
narrowing of the smaller blood vessels is observed, and suddenly the
narrowed vessels develop a peculiar pattern consisting of alternating
regions of constrictions and dilatations, giving the vessels the
appearance of sausages on a string (Fig.~1). The `sausage-string'
pattern may cause severe damages to the blood vessels because plasma
and macromolecules are transported into the vessel wall in the dilated
regions. The sausage-string pattern has been observed in small vessels
from many organs, including the brain, the gut, and the kidney
\cite{finng}.

Despite several decades of research, the mechanism causing the
`sausage-string' pattern has remained unknown \cite{finng}. It has been
suggested that it represents a `blow out' of the vessel wall due to
the high blood pressure \cite{bg}, but this seems unlikely for several
reasons. The sausage-string pattern occurs in the smaller
vessels (small arteries and large arterioles), and here the pressure
elevation is relatively small compared to that in the larger
arteries. Secondly, the phenomenon is highly reproducible
\cite{jg}. If the infusion of the vasoconstricting agent is stopped,
the normal, uniform `cylindrical' structure is restored. Restoring the
infusion causes again an extreme, uniform vasoconstriction followed by
the reappearance of the sausage-string pattern. A third spectacular feature
of the phenomenon is its periodicity with constrictions and
dilatations occurring in a regular and repetitive pattern.

In this Letter we present a simple anisotropic, elastic model of the
vessel wall. We show that under certain hypertensive conditions an
instability occurs which leads to a periodic pattern of constrictions
and dilatations along the vessel. Our theory provides predictions for
the conditions under which the cylindrical form of a blood vessel
becomes unstable.

To be specific, a cylindrical shaped blood vessel is unstable if a
small axial symmetric perturbation of the inner radius, $r\to r+u(z)$,
grows (Fig.~2). To determine the stability, we must therefore know
the dynamic equation for the perturbation $u(z,t)$. To this end, we
invoke the continuity equation, $\partial_t(\pi r^2)=-\partial_zJ$,
associating a local change of the cross-sectional area at a downstream
site $z$ with a fluid flux $J(z)$. The flux is related to the
transmural pressure $P$, by $J=-c(r)\partial_zP$, where $c(r)$ is the
vascular conductance \cite{hpa}. From the continuity equation and the
flux-pressure relation, the dynamic equation to lowest order in the
perturbation follows,
\begin{equation} \label{drp}
\partial_t u=\frac{c(r)}{2\pi r}\partial_z^2P~.
\end{equation}

As a simple illustration, consider first a very thin vessel wall, for
which the pressure is given by the Laplace form \cite{Fung}
$P=(T/R)+(T_z/R_z)$, where $T$ and $T_z$ are the tensions
circumferential to and parallel with the cylinder axis $z$, and $1/R$
and $1/R_z$ are the curvatures in the corresponding directions
\cite{rad}. Assume furthermore that the tensions are constant and
identical, $T_z=T$. Inserting the above expression for the pressure
into Eq.~(\ref{drp}) and retaining only first order terms in $u$, we
get
\begin{equation} \label{ekp}
\partial_t u = -\frac{Tc(r)}{2\pi r^3}
[\partial_z^2 u + r^2\partial_z^4 u]~.
\end{equation}
For a given periodic perturbation, $u=u_k(t)\cos(kz)$, we have
$u_k(t) \sim u_k(0)e^{\lambda_kt}$, where
\begin{equation} \label{elk}
\lambda_k=\frac{Tc(r)}{2\pi r^3}\; k^2[1-r^2k^2]~.
\end{equation}
Thus the vessel wall is unstable to modes with $rk<1$. The dominant
mode, where $\lambda_k$ is maximal, is at $k=1/(\sqrt{2}r)$.

The above instability is the well-known Rayleigh instability
\cite{Pla,Ray}. The theory explains why a cylindrical column of water
with surface tension $T$ is unstable at all radii. However,
cylindrical structures may be stable due to a reluctance against
bending \cite{ng}. Still an instability may occur if the tension $T$
can be brought to exceed a critical value of order $\kappa/r^2$,
$\kappa$ being the bending modulus \cite{ng}. This is demonstrated by
the so-called `pearling' instability, recently observed by Bar-Ziv and
Moses \cite{BZM} in tubular lipid membranes.

For blood vessels the width of the vessel wall cannot be neglected.
Furthermore, the stress is highly nonlinear and strongly dependent on
the strain \cite{Fung}. Taking the width $w$ of the blood vessel into
account, the Laplacian form for the pressure is replaced by an
integral,
\begin{equation} \label{lpi}
P=\int_{r}^{r+w}[\; S\frac{1}{\tilde{r}[1+(\partial_z\tilde{r})^2]^{1/2}}
-S_z\frac{\partial_z^2\tilde{r}}{[1+(\partial_z\tilde{r})^2]^{3/2}}
\; ]\; d\tilde{r}~,
\end{equation}
where $S$ is the angular stress, and $S_z$ is the stress along the
vessel. The stresses, defined as the forces per actual cross-sectional
area, are related to the experimentally measured idealized stresses
$\sigma$ and $\sigma_z$, defined as the forces per relaxed
cross-sectional area, $S=\gamma\gamma_z\sigma$ and
$S_z=\gamma\gamma_z\sigma_z$ \cite{FCH}. Here $\gamma$ and $\gamma_z$
are the normalized lengths \cite{nlen} in the angular and vessel
direction. Since the length of a vessel remains almost constant during
a contraction, $\gamma_z$ is here assumed to be constant,
$\gamma_z=\gamma_0$. Correspondingly, the stress $\sigma_z$ is
replaced by a constant $\sigma_0$.
We note that the width $w$ of the vessel wall changes when the inner
radius $r$ changes (Fig.~2). Assuming that the cross-sectional area
of the vessel wall is constant, the radius dependence of $w$ is given,
when the inner radius $\rho$ and wall thickness $\omega$ are known for
the angularly relaxed state ($\gamma=1$) \cite{area}.

For small perturbations, the relevant expression for the pressure
reduces to
\begin{eqnarray} \label{psi}
P=\gamma_0\int_{\rho}^{\rho+\omega}\;
[\sigma-\sigma_0r\partial_z^2r][\tilde{\rho}^2-\rho^2+r^2]^{-1/2}\;
d\tilde{\rho}~.
\end{eqnarray}
For the angular direction, the stress $\sigma$ depends on the
normalized length
$\gamma=[\tilde{\rho}^2-\rho^2+r^2]^{1/2}/\tilde{\rho}$. To first
order in the perturbation $u(z,t)$, we find
\begin{equation} \label{P}
P=P_0(r)+I(r)u-I_0(r)\partial_z^2u~,
\end{equation}
where
\begin{eqnarray} \label{P0}
P_0(r)&=&\gamma_0\int_{\rho}^{\rho+\omega}\;
\sigma[\tilde{\rho}^2-\rho^2+r^2]^{-1/2}\;
d\tilde{\rho}~,\\
I_0(r)&=&\gamma_0\sigma_0r\int_{\rho}^{\rho+\omega}\;
[\tilde{\rho}^2-\rho^2+r^2]^{-1/2}\;
d\tilde{\rho}\cr
&=&\gamma_0\sigma_0r\log[1+(\omega+w)/(\rho+r)]~,
\end{eqnarray}
and
\begin{eqnarray} \label{Ir}
I(r)=\frac{d}{dr}P_0(r)=\gamma_0\int_{\rho}^{\rho+\omega}\;
\tilde{\rho}^{-1}\frac{d}{d\gamma}\left[\frac{\sigma}{\gamma}\right]\;
\frac{\partial\gamma}{\partial r}\;
d\tilde{\rho}~.
\end{eqnarray}
The partial derivatives of $\gamma$ with respect to $r$ and $\tilde{\rho}$
are related, $\tilde{\rho}^{-1}(\partial\gamma/\partial r)=
r(\rho^2-r^2)^{-1}(\partial\gamma/\partial\tilde{\rho})$,
and $I(r)$ can be expressed in terms of the normalized length $\gamma$,
\begin{equation} \label{p1i}
I(r)=\frac{\gamma_0\gamma_r}{\rho(1-\gamma_r^2)}\left[
\frac{\sigma(\gamma_w)}{\gamma_w}-\frac{\sigma(\gamma_r)}{\gamma_r}\right]~,
\end{equation}
where $\gamma_r=r/\rho$ is the normalized inner radius, and
$\gamma_w=(r+w)/(\rho+\omega)$ is the normalized outer radius. We note
that $I(r)$ is not singular at $\gamma_r=1$, where also $\gamma_w=1$.
Inserting Eq.~(\ref{P}) into Eq.~(\ref{drp}) we get for a given
periodic perturbation, $u = u_k(t)\cos(kz)$, that $u_k(t)\sim
u_k(0)e^{\lambda_kt}$, where
\begin{equation} \label{lke}
\lambda_k=\frac{c(r)}{2\pi r}\; k^2[-I(r)-I_0(r)k^2]~.
\end{equation}
The value of $I_0(r)$ is always positive. Thus, it is the sign of
$I(r)$ that determines the stability of the vessel wall. If $I(r)$ is
positive the cylindrical shape is stable for all modes. If $I(r)$ is
negative, the cylindrical shape is unstable to modes with
$k^2<|I|/I_0$.

As seen from the expression for $I$, Eq.~(\ref{p1i}), the important
quantity is $\sigma/\gamma$. The point of instability is where
$\sigma/\gamma$ calculated at the inner radius equals the value of
$\sigma/\gamma$ at the outer radius.  This can be illustrated
geometrically by drawing a line in the plot of $\sigma$ versus
$\gamma$ (Fig.~3) from $(0,0)$ through
$(\gamma_r,\sigma(\gamma_r))$. If the point
$(\gamma_w,\sigma(\gamma_w))$ lies above this line, $I(r)$ is positive
and the cylindrical form is stable. If however the point
$(\gamma_w,\sigma(\gamma_w))$ lies below this line, $I(r)$ is
negative, leading to an instability of the cylindrical form.

Under normal physiological conditions, the angular stress $\sigma$ in
blood vessels increases linearly to exponentially with the normalized
length \cite{Fung,gd} (Fig.~3), and the value of $I(r)$ is therefore
positive (Fig.~4). This ensures that the blood vessel keeps its
cylindrical shape -- no bending arguments are needed to explain the
stability of the cylindrical shape of a blood vessel. However, when
acute hypertension is induced by infusion of a strong vasoconstricting
agent like angiotensin II, there will be a substantial reduction of
the inner radius in small arteries and large arterioles due to
contraction of the smooth muscle cells. The operating point for the
vessel will now be on a less steep part of the $\sigma-\gamma$ curve
(Fig.~3), and when the radius is reduced below a certain inner radius
$r_c$ at which $\sigma(\gamma_w)/\gamma_w=\sigma(\gamma_r)/\gamma_r$,
the value of $I(r)$ becomes negative (Fig.~4). This will result in an
instability of the cylindrical form, giving rise to the
`sausage-string' pattern. The dominant (fastest growing) mode is given
by $k=[|I|/(2I_0)]^{1/2}$, which will correspond to `sausages' of
length $\ell=2\pi\sqrt{2I_0/|I|}$. Insertion of typical values
\cite{typ} for the various parameters of the model yields $\ell\approx
2\pi\rho$, hence the length of the `sausages' will be 5-10 times the
radius of the relaxed vessel. The theory therefore predicts that the
`sausages' will have an elongated shape with a length that decreases
as the vessel radius gets smaller. This is in good agreement with
experimental observations \cite{jg} (Fig.~1).

A way to view the basic phenomenon underlying the instability is to
note that, when $I(r)$ becomes negative, the pressure at slightly
larger radii is smaller than at slightly smaller radii. Accordingly,
the resulting flow $J$ will be directed from low-radii regions to
high-radii regions, causing the small radii to become even smaller,
and the large radii to become larger. This continues until the
pressure stabilizes at a value which is the same for both the large
radius $r_{max}$ and the small radius $r_{min}$. The stabilization is
only possible, because the pressure for radii above the instability,
$r>r_c$, again increases with $r$. The theory allows an estimate of
the radius in the dilated regions, $r_{max}$. Assuming that
$r_{min}/\rho$ is small (close to zero), we can estimate the final
value of $r_{max}$ by the condition $P(r_{max})=P(0)$. Interestingly,
the almost linear stress function in the region above $r_c$ (Fig.~3)
gives rise to a decay of $I(r)$ in the same region (Fig.~4). As a
consequence $r_{max}$ can become quite large. However, close to
$\gamma = 1$, the stress increases exponentially due to the elastic
properties of the vessel wall \cite{Fung}, and the value of $I(r)$
will increase rapidly. This will effectively prevent $r_{max}$ from
attaining a value substantially larger than the relaxed radius,
$\rho$, of the vessel. However, $r_{max}$ may be larger than the
working radius of the vessel under normal physiological conditions,
because the normal working radius is smaller than the relaxed radius
\cite{Fung}. This may explain why previous work have suggested that
the dilated regions represented a `blow out' due to mechanical failure
of the vessel wall \cite{bg}.

The `sausage-string' pattern following infusion of angiotensin II have
been found to occur predominantly in small arteries and large
arterioles \cite{jg}. The present analysis predicts that large vessels
will be stable. Their operating point are on the steep portion of the
$\sigma-\gamma$ curve due to their high pressure. As seen from Fig.~3,
the contraction is here limited, thus preventing the larger vessels
from reducing their radius below the critical value $r_c$. As arterial
vessels gets smaller the wall-to-lumen ratio $\omega/\rho$ increases
\cite{Fung}. From the expression for $I(r)$, Eq.~(\ref{p1i}), we find
that $r_c$ as well as $r_{max}$ decreases with increasing
wall-to-lumen ratio. Hence, the `sausage-string' instability is less
likely to appear in blood vessels with large wall-to-lumen ratios. It
seems that the transmural pressure and the contractile potential sets
an upper limit, and the wall-to-lumen ratio a lower limit for vessels
that will undergo the `sausage-string' instability in response to an
acute increase in blood pressure.

In summary, we have demonstrated that during severe vasoconstriction,
the normal cylindrical shape of a blood vessel may become unstable,
and as a result the vessel exhibit a periodic pattern of constrictions
and dilatations. The sausage-string pattern is not caused by a mechanical
failure of the vessel wall due to the high blood pressure, but is the
expression of an instability. The instability is related to the
Rayleigh instability, and to the `pearling' instability seen in tubular
lipid membranes. The mechanism behind the instability, however, is
novel, involving the nonlinear elasticity characteristics of the
vessel wall. The developed theory explains many of the key features
observed experimentally, e.g.\ the predominant occurrence in small
arteries and large arterioles, and most likely in those with small
wall-to-lumen ratios.

The present study was supported by grants from the Danish Natural
Science Research Council, the Danish Medical Research Council, the
Novo-Nordisk Foundation and the Danish Heart Association.

\begin{figure}
\vskip 4mm
\caption{
{\em In vivo} micrograph of rat intestinal arterioles showing a
typical `sausage-string' pattern following an acute increase in
blood pressure induced by intravenous infusion of angiotensin II. The
neighboring vessels not showing constrictions and dilatations are the
corresponding venules. From [4]
with permission.}
\end{figure}

\begin{figure}
\centerline{\psfig{file=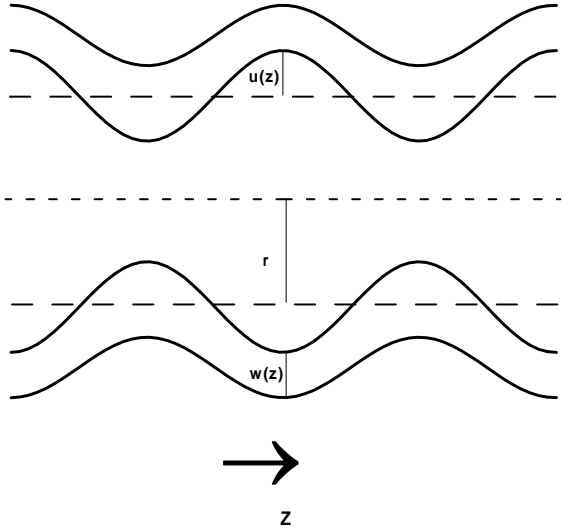,width=0.45\textwidth}}
\vskip 4mm
\caption{
A schematic picture of a blood vessel of inner radius $r$ undergoing a
perturbation $u(z)$. The wall thickness $w(z)$ is larger at smaller
radii since the circumference is smaller.}
\end{figure}

\begin{figure}
\centerline{\psfig{file=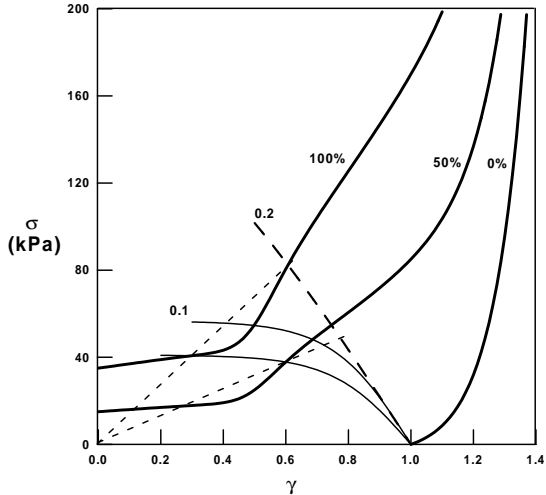,width=0.4\textwidth}}
\vskip 4mm
\caption{
A schematic plot of typical stress-strain relations for arterioles
(adapted from [16]).
The three heavy solid curves correspond to a completely relaxed vessel
(0\%), a vessel where the smooth muscle cells are half maximally
activated (50\%), and a vessel where the smooth muscle cells are
maximally activated (100\%). The thin solid lines show how the points
$(\gamma_r,\sigma(\gamma_r))$ and $(\gamma_w,\sigma(\gamma_w))$
[marked 0.1] move with muscle cell activation for an arteriole with
wall-to-lumen ratio $\omega/\rho$ = 0.1. The point of instability
($r=r_c$) for the cylindrical form of the blood vessel is illustrated
geometrically by thin dashed lines from $(0,0)$ through
$(\gamma_r,\sigma(\gamma_r))$. The instability point is where
$\sigma(\gamma_w)/\gamma_w$ equals $\sigma(\gamma_r)/\gamma_r$. The
thick dashed line [marked 0.2] shows how the point
$(\gamma_w,\sigma(\gamma_w))$ move with muscle cell activation for an
arteriole with $\omega/\rho$ = 0.2, keeping the same curve for
$(\gamma_r,\sigma(\gamma_r))$.}
\end{figure}

\begin{figure}
\centerline{\psfig{file=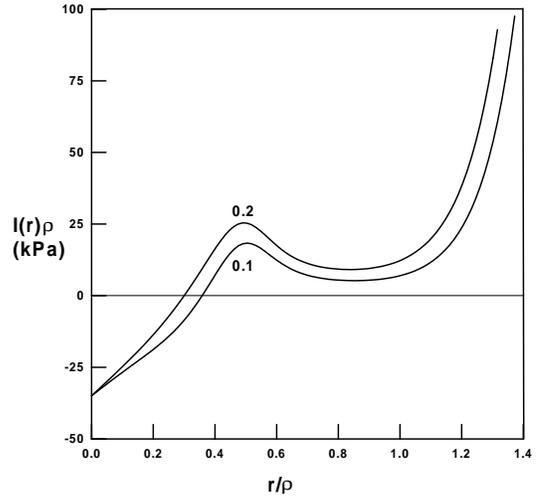,width=0.4\textwidth}}
\vskip 4mm
\caption{
A plot of the stability measure $I(r)$ at large muscle cell
activation for two different wall-to-lumen ratios, $\omega/\rho=0.1$
and $\omega/\rho=0.2$. The cylindrical form of a blood vessel becomes
unstable when $I$ becomes negative. An almost linear stress-strain
relation in a region above $r_c$ gives rise to a decay of $I$. Above
$\gamma=1$, where the stress increases exponentially, also $I(r)$
increases exponentially.}
\end{figure}


\begin{thebibliography}{99}
\bibitem{hpa}
In the Hagen-Poiseuille approximation, the fluid conductance is
$c(r)=\pi r^4/(8\eta)$, where $\eta$ is the dynamic viscosity of the
fluid (blood), but the specific form of $c(r)$ is not crucial for our
purpose. Moreover, we neglect the pressure drop along the vessel,
noting that this is much smaller than the transmural pressure.
\bibitem{rad}
The principal curvatures $1/R$ and $1/R_z$ are given by the two
relations:
\begin{eqnarray}
\frac{1}{R}=\frac{1}{r[1+(\partial_zr)^2]^{1/2}}~, \: \mbox{ and }
\: \frac{1}{R_z}=\frac{-\partial_z^2r}{[1+(\partial_zr)^2]^{3/2}}~.
\nonumber
\end{eqnarray}
\bibitem{nlen}
The normalized length, $\gamma$, is equal to $L/L_0$ where $L$ is the
actual length of a tissue strip and $L_0$ is the resting length. The
corresponding strain, $\epsilon$, is by definition equal to $\gamma -
1$.
\bibitem{area}
We have $(r+w)^2-r^2=(\rho+\omega)^2-\rho^2$. This equation suggests a
useful change of variable, from $\tilde{r}$ to $\tilde{\rho}$, where
$\tilde{r}^2-r^2=\tilde{\rho}^2-\rho^2$.  While $\tilde{r}$ varies
between values $r$ and $r+w$, which under perturbations changes along
the $z$ axis, $\tilde{\rho}$ varies between the fixed values $\rho$
and $\rho+\omega$. The normalized length $\gamma$ at a radius
$\tilde{r}$ is simply the ratio between $\tilde{r}$ and its relaxed
value $\tilde{\rho}$, i.e.\ $\gamma=\tilde{r}/\tilde{\rho}$.
\bibitem{typ}
For $\omega/\rho$ = 0.1, and $\sigma_0$ = 100~kPa, we have
$|I|\rho\approx 2I_0/\rho\approx 10$~kPa.
\end{thebibliography}
\end{document}